\documentclass[final,5p,twocolumn,feyn,feynmf]{elsarticle}
\usepackage[utf8]{inputenc}
\usepackage{xcolor} 
\usepackage{mathrsfs}
\usepackage{subcaption}
\usepackage{adjustbox}
\usepackage{graphicx}% Include figure files
\usepackage{dcolumn}% Align table columns on decimal point
\usepackage{bm}% bold math
\usepackage[T1]{fontenc}
\usepackage{booktabs, array, mathptmx, float, tabularx, booktabs, lipsum, amsmath,multirow}
\usepackage{siunitx, xcolor}
\usepackage[version=4]{mhchem}
\usepackage[colorlinks,linkcolor=blue,anchorcolor=blue,citecolor=blue]{hyperref}
\journal{Physics Letters B}

\begin{document}
\begin{frontmatter}
\title{Constraints on the axial-vector and pseudo-scalar mediated WIMP-nucleus interactions from PandaX-4T experiment}
% !TEX root = ../main.
\address[shKeyLab]{School of Physics and Astronomy, Shanghai Jiao Tong University, Key Laboratory for Particle Astrophysics and Cosmology (MoE), Shanghai Key Laboratory for Particle Physics and Cosmology, Shanghai 200240, China}
\address[BUAA]{School of Physics, Beihang University, Beijing 102206, China}
\address[BUAALab]{Beijing Key Laboratory of Advanced Nuclear Materials and Physics, Beihang University, Beijing, 102206, China}
\address[zzu]{School of Physics and Microelectronics, Zhengzhou University, Zhengzhou, Henan 450001, China}
\address[USTClab]{State Key Laboratory of Particle Detection and Electronics, University of Science and Technology of China, Hefei 230026, China}
\address[USTCdep]{Department of Modern Physics, University of Science and Technology of China, Hefei 230026, China}
\address[pku]{School of Physics, Peking University, Beijing 100871, China}
\address[YaLongSD]{Yalong River Hydropower Development Company, Ltd., 288 Shuanglin Road, Chengdu 610051, China}
\address[CHEPpku]{Center for High Energy Physics, Peking University, Beijing 100871, China}
\address[SDUdep]{Research Center for Particle Science and Technology, Institute of Frontier and Interdisciplinary Scienc, Shandong University, Qingdao 266237, Shandong, China}
\address[SDUlab]{Key Laboratory of Particle Physics and Particle Irradiation of Ministry of Education, Shandong University, Qingdao 266237, Shandong, China}
\address[UMD]{Department of Physics, University of Maryland, College Park, Maryland 20742, USA}
\address[TDLee]{Tsung-Dao Lee Institute, Shanghai Jiao Tong University, Shanghai, 200240, China}
\address[MESJTU]{School of Mechanical Engineering, Shanghai Jiao Tong University, Shanghai 200240, China}
\address[SYU]{School of Physics, Sun Yat-Sen University, Guangzhou 510275, China}
\address[SYUSFI]{Sino-French Institute of Nuclear Engineering and Technology, Sun Yat-Sen University, Zhuhai, 519082, China}
\address[NKU]{School of Physics, Nankai University, Tianjin 300071, China}
\address[FDU]{Key Laboratory of Nuclear Physics and Ion-beam Application (MOE), Institute of Modern Physics, Fudan University, Shanghai 200433, China}
\address[USST]{School of Medical Instrument and Food Engineering, University of Shanghai for Science and Technology, Shanghai 200093, China}
\address[SJTUSC]{Shanghai Jiao Tong University Sichuan Research Institute, Chengdu 610213, China}
\address[SPEIT]{SJTU Paris Elite Institute of Technology, Shanghai Jiao Tong University, Shanghai, 200240, China}

\author[shKeyLab]{Zhou Huang}
\author[shKeyLab]{Chencheng Han}
\author[shKeyLab]{Abdusalam Abdukerim}
\author[shKeyLab]{Zihao Bo}
\author[shKeyLab]{Wei Chen}
\author[shKeyLab,SJTUSC]{Xun Chen}
\author[YaLongSD]{Yunhua Chen}
\author[SYU]{Chen Cheng}
\author[SYUSFI]{Zhaokan Cheng}
\author[TDLee]{Xiangyi Cui}
\author[NKU]{Yingjie Fan}
\author[FDU]{Deqing Fang}
\author[FDU]{Changbo Fu}\
\author[pku]{Mengting Fu}
\author[BUAA,BUAALab,zzu]{Lisheng Geng}
\author[shKeyLab]{Karl Giboni}
\author[shKeyLab]{Linhui Gu}
\author[YaLongSD]{Xuyuan Guo}
\author[shKeyLab]{Ke Han}
\author[shKeyLab]{Changda He}
\author[YaLongSD]{Jinrong He}
\author[shKeyLab]{Di Huang}
\author[USST]{Yanlin Huang}
\author[SJTUSC]{Ruquan Hou}
\author[UMD]{Xiangdong Ji}
\author[MESJTU]{Yonglin Ju}
\author[shKeyLab]{Chenxiang Li}
\author[SYU]{Jiafu Li}
\author[YaLongSD]{Mingchuan Li}
\author[MESJTU]{Shu Li}
\author[TDLee]{Shuaijie Li}
\author[USTClab,USTCdep]{Qing Lin}
\author[shKeyLab,TDLee,SJTUSC]{Jianglai Liu\fnref{fn1}}
\ead{jianglai.liu@sjtu.edu.cn}
\author[SDUdep,SDUlab]{Xiaoying Lu}
\author[pku]{Lingyin Luo}
\author[USTCdep]{Yunyang Luo}
\author[shKeyLab]{Wenbo Ma}
\author[FDU]{Yugang Ma}
\author[pku]{Yajun Mao}
\author[shKeyLab,SJTUSC]{Yue Meng}
\author[shKeyLab]{Xuyang Ning}
\author[YaLongSD]{Ningchun Qi}
\author[shKeyLab]{Zhicheng Qian}
\author[SDUdep,SDUlab]{Xiangxiang Ren}
\author[SDUdep,SDUlab]{Nasir Shaheed}
\author[YaLongSD]{Changsong Shang}
\author[shKeyLab]{Xiaofeng Shang}
\author[BUAA]{Guofang Shen}
\author[shKeyLab]{Lin Si}
\author[YaLongSD]{Wenliang Sun}
\author[UMD]{Andi Tan}
\author[shKeyLab,SJTUSC]{Yi Tao}
\author[SDUdep,SDUlab]{Anqing Wang}
\author[SDUdep,SDUlab]{Meng Wang}
\author[FDU]{Qiuhong Wang}
\author[shKeyLab, SPEIT]{Shaobo Wang}
\author[pku]{Siguang Wang}
\author[SYU]{Wei Wang}
\author[MESJTU]{Xiuli Wang}
\author[shKeyLab,SJTUSC,TDLee]{Zhou Wang}
\author[SYUSFI]{Yuehuan Wei}
\author[SYU]{Mengmeng Wu}
\author[shKeyLab]{Weihao Wu}
\author[shKeyLab]{Jingkai Xia}
\author[UMD]{Mengjiao Xiao}
\author[SYU]{Xiang Xiao}
\author[TDLee]{Pengwei Xie}
\author[shKeyLab,SJTUSC]{Binbin Yan}
\author[USST]{Xiyu Yan}
\author[shKeyLab]{Jijun Yang}
\author[shKeyLab]{Yong Yang}
\author[NKU]{Chunxu Yu}
\author[SDUdep,SDUlab]{Jumin Yuan}
\author[FDU]{Zhe Yuan}
\author[shKeyLab]{Xinning Zeng}
\author[UMD]{Dan Zhang}
\author[shKeyLab]{Minzhen Zhang}
\author[YaLongSD]{Peng Zhang}
\author[shKeyLab]{Shibo Zhang}
\author[SYU]{Shu Zhang}
\author[shKeyLab]{Tao Zhang}
\author[SDUdep,SDUlab]{Yingxin Zhang}
\author[TDLee]{Yuanyuan Zhang}
\author[shKeyLab]{Li Zhao}
\author[USST]{Qibin Zheng}
\author[YaLongSD]{Jifang Zhou}
\author[shKeyLab,SJTUSC]{Ning Zhou\corref{cor1}}
\ead{nzhou@sjtu.edu.cn}
\author[BUAA]{Xiaopeng Zhou}
\author[YaLongSD]{Yong Zhou}
\author[shKeyLab]{Yubo Zhou}
\fntext[fn1]{Spokesperson of PandaX Collaboration}
\cortext[cor1]{Corresponding author}

\begin{abstract}
We present the constraints on the axial-vector and pseudo-scalar mediated WIMP-nucleus interactions 
from the PandaX-4T experiment, using the data set corresponding to a total exposure of 0.63~tonne$\cdot$year. 
No significant signal excess is observed, and the most stringent constraints to date on the mediator mass 
are set for the axial-vector and pseudo-scalar simplified models. The maximum excluded mass of an axial-vector mediator is $825~\si{GeV}/c^2$ and that of a pseudo-scalar mediator is $106~\si{GeV}/c^2$. In addition, the upper limits on the spin-dependent WIMP-nucleon scattering cross-section for conventional neutron-only and proton-only interactions 
are derived.
\end{abstract}
\end{frontmatter}

There is plenty of evidence indicating that a large amount of dark matter exists in our universe. However, its nature is still elusive~\cite{Bertone:2004pz}.
Weakly interacting massive particle (WIMP) is a plausible 
dark matter candidate which arises naturally from many theories beyond the Standard Model of particle physics like the super-symmetry. Various methods are utilized to hunt for WIMPs, including direct search, indirect search, and collider search. 
Based on the dual-phase xenon time projection chamber (TPC), the sensitivity of WIMP-nucleus interaction has improved significantly over the past decades ~\cite{LUX2016,pandax1_full_exposure,pandax2_final,PandaX-4T-SI,XENON100_2016,XENON-SI-2018}.

The PandaX-4T experiment, located at the China Jinping Underground Laboratory (CJPL)~\cite{CJPL_intro,CJPL2_intro}, 
is one of the largest xenon TPC experiments with 3.7-tonne of liquid xenon in the sensitive volume.
The cylindrical TPC is covered by 24 polytetrafluoroethylene (PTFE) panels on the side 
and 368 Hamamatsu R11410-23 3-inch photomultiplier tubes (PMTs) on the top and bottom, with four electrodes placed inside to construct the drift and extraction electric field.
Once an incoming energetic particle scatters off xenon nuclei or electrons, prompt scintillation light ($S1$) and ionized electrons are generated, the latter then drift upward to the liquid xenon surface and produce the electroluminescence light ($S2$) in the gaseous xenon region under the electric fields. 
Both $S1$ and $S2$ are detected by the PMTs, from which the scattering position and deposited energy can be reconstructed. The ratio of $S2$ to $S1$ provides powerful signal-background discrimination. A more detailed description of the PandaX-4T experiment can be found in Refs.~\cite{PandaX:2018wtu, P4_neutron_bkg,PandaX-4T-SI}. 

The PandaX-4T experiment has set stringent constraints on the spin-independent (SI) WIMP-nucleon cross-section upper limits, excluding $3.8\times 10^{-47}\rm cm^2$ for a WIMP mass of 40 GeV$/c^2$ at 90\% confidence level (C.L.)~\cite{PandaX-4T-SI}, \textcolor{black}{
which is based on an effective field theory (EFT) describing the Dirac fermion WIMP particle interaction with the nucleon. The physics behind the EFT can be revealed from a simplified model, where
the interaction is introduced by a mediator with couplings to WIMPs and quarks. A vector or scalar mediator produces the SI WIMP-nucleon interaction.
For an axial-vector and pseudo-scalar mediator, the tree-level process yields the spin-dependent (SD) WIMP-nucleus interaction instead~\cite{Bell:2018zra,Alanne:2022eem}. Xenon-based experiments are also sensitive to the SD interactions thanks to the significant abundance of odd-$A$ xenon isotopes with non-zero spin (26.4\% spin-1/2 $\rm ^{129}Xe$ and 21.2\% spin-3/2 $\rm ^{131}Xe$ in natural xenon)~\cite{PandaX-II-EFT, XENON1T-SD}.  The XENON1T experiment has pushed down the SD neutron-only cross-section limits to $6.3\times 10^{-42} \rm cm^2$ for WIMP mass of 30 GeV$/c^2$. As for pseudo-scalar mediator, the tree-level SD process has a strong momentum suppression~\cite{Li:2018qip,Li:2019fnn}, there are no effective searches for this pseudo-scalar mediated SD process in direct detection experiments. However, under the simplified model, the next-to-leading-order (NLO) loop-level process does not have this suppression and yields a SI process, which allows the direct detection to be able to probe the pseudo-scalar mediator too. In this letter, we present the search results of SD WIMP-nucleus interactions and constraints on the axial-vector and pseudo-scalar mediators, based on the PandaX-4T commissioning run data.} 

As for axial-vector couplings to quarks and WIMPs, the WIMP-nucleus SD cross-section can be evaluated from chiral effective field theory (EFT)~\cite{Klos:2013rwa,Bishara:2017pfq,Bishara:2017nnn}, which is related to the total spin expectation values of the protons and neutrons in the nucleus in the limit of zero momentum transfer. Therefore, a common practice is to consider two special cases where the WIMPs couple only to protons or to neutrons. The corresponding scattering cross-section is expressed as below,
\begin{equation}
	\sigma_A = \frac{4 \pi }{3(2J+1)}\left(\frac{\mu_A}{\mu_N}\right)^2 S_N(q^2)\sigma_N,
\end{equation}
where $N$ denotes the proton ($p$) or neutron ($n$),
$\mu_N$ is the reduced mass for WIMP scattering off the nucleon,$\mu_A$ is the reduced mass of WIMP and the target nucleus,
$S_N(q^2)$ is the structure factor for ``proton-only" or ``neutron-only'' cases~\cite{Klos:2013rwa}. Due to the cancellation between spins of nucleon pairs, the SD cross-section with a nucleus is dominated by the unpaired nucleon, which is neutron for $\rm ^{129}Xe$ and $\rm ^{131}Xe$. It should be noticed that the cross-section of the ``proton-only" case is significantly enhanced due to chiral two-body currents involving the exchange of a pion.

\begin{figure}[htbp]
\centering
\includegraphics[width=0.9\columnwidth]{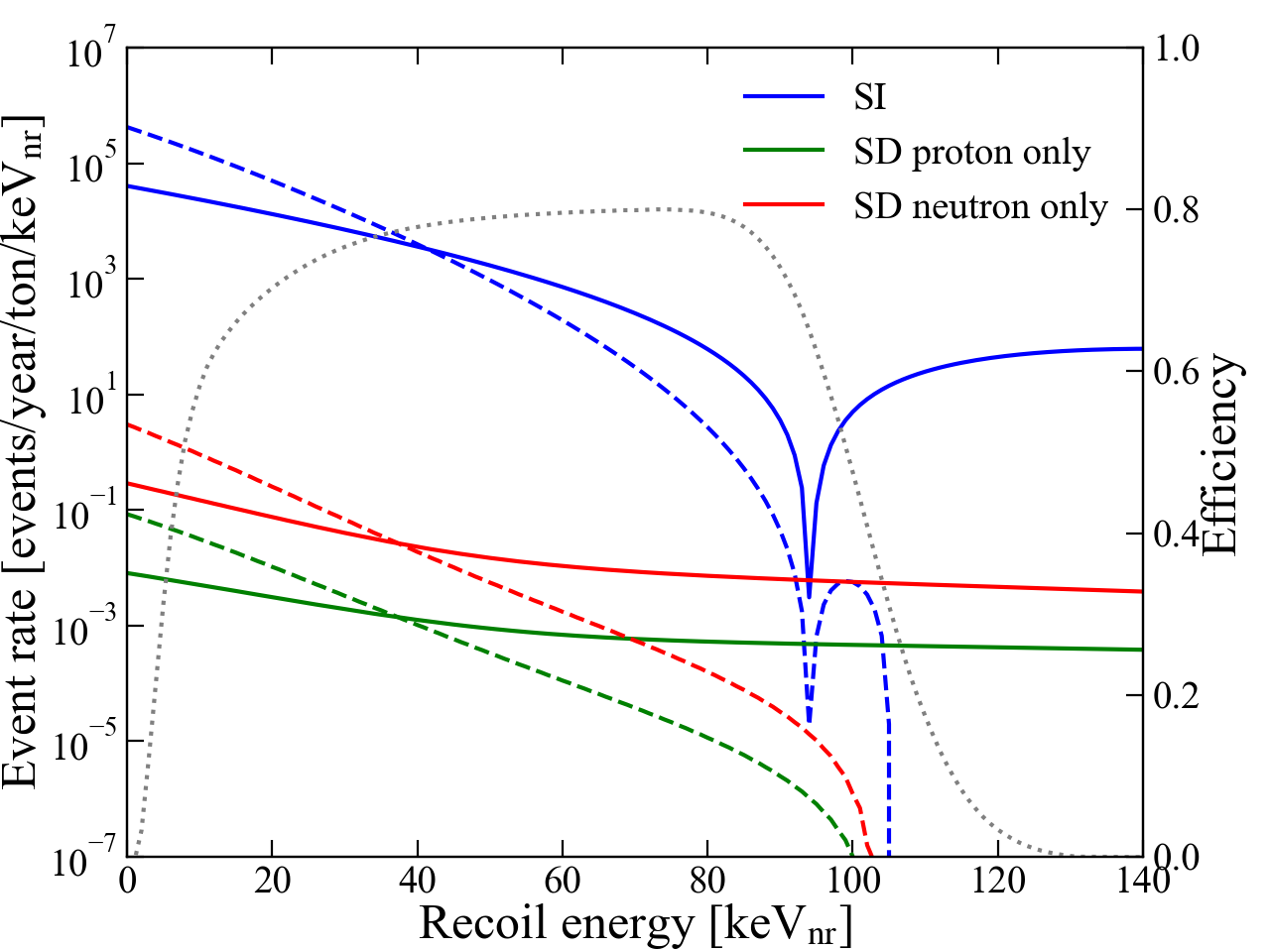}
\caption{Nuclear recoil energy spectra for the scattering of WIMP on PandaX-4T detector.
		 The SI, SD proton-only, and SD neutron-only are illustrated in blue, green
		 and red curves with $\sigma = 10^{-41}$cm$^2$ respectively. The dashed and solid curves represent the 
		 40~GeV/c$^2$ and 400~GeV/c$^2$ WIMPs respectively.
		 The detection efficiency of the PandaX-4T experiment is plotted in a dotted curve
		 with the corresponding axis on the right.}
\label{fig:Espectrum}
\end{figure}

The differential event rate for the WIMP SD interaction 
can be then written as~\cite{Savage:2008er,Lewin:1995rx}
\begin{equation}
	\frac{dR}{dE} = \frac{\rho_{\chi}\sigma_A}{2m_{\chi}\mu_A^2}\eta(E),
\end{equation}
where $m_\chi$ and $\rho_\chi$ are the WIMP mass and local density respectively,
$\sigma_A$ is the WIMP-nucleus cross-section,
$\eta$ is the mean inverse speed of WIMP velocity distribution.
It is recommended that $\rho_\chi$ = 0.3~GeV/(c$^2\cdot$cm$^3$)
and the WIMP motion obeys the standard halo model~\cite{Baxter:2021pqo}.
Explicitly, 
the most probable velocity of WIMP in Maxwellian distribution $v_0$ = 238~km/s,
the galactic escape velocity $v_{\rm esc}$ = 544~km/s
and the Earth velocity $v_{\rm E}$ = 254~km/s are adopted in this analysis.
The typical event rate as a function of the nuclear recoil energy is shown in Fig.~\ref{fig:Espectrum}.
Both the SD ``proton-only" and ``neutron-only" recoil energy spectra are plotted in addition to the SI case for the WIMP mass of 40~GeV/c$^2$ and 400~GeV/c$^2$,
with the detection efficiency~\cite{PandaX-4T-SI} overlaid.

%\section{Event selection}
We use the same data set and selections as the WIMP SI analysis~\cite{PandaX-4T-SI}. In brief, the $S1$ signal region of interest requires the $S1$ in the range of 2 to 135~photoelectron(PE) in the fiducial volume, and the $S2$ signal collected in the bottom PMT array ($S2_{\rm b}$) is adopted to avoid bias from PMT saturation. 
The data is divided into five subsets, according to the detector status.
The background models also inherit from Ref.~\cite{PandaX-4T-SI}
and the detailed values can be found in the Table~II of Ref.~\cite{PandaX-4T-SI}.
The total live time is 86.0 days and the resulting exposure is 0.63 tonne$\cdot$year.

%\section{Limits}

\begin{figure}[htbp]
\centering
\includegraphics[width=0.9\columnwidth]{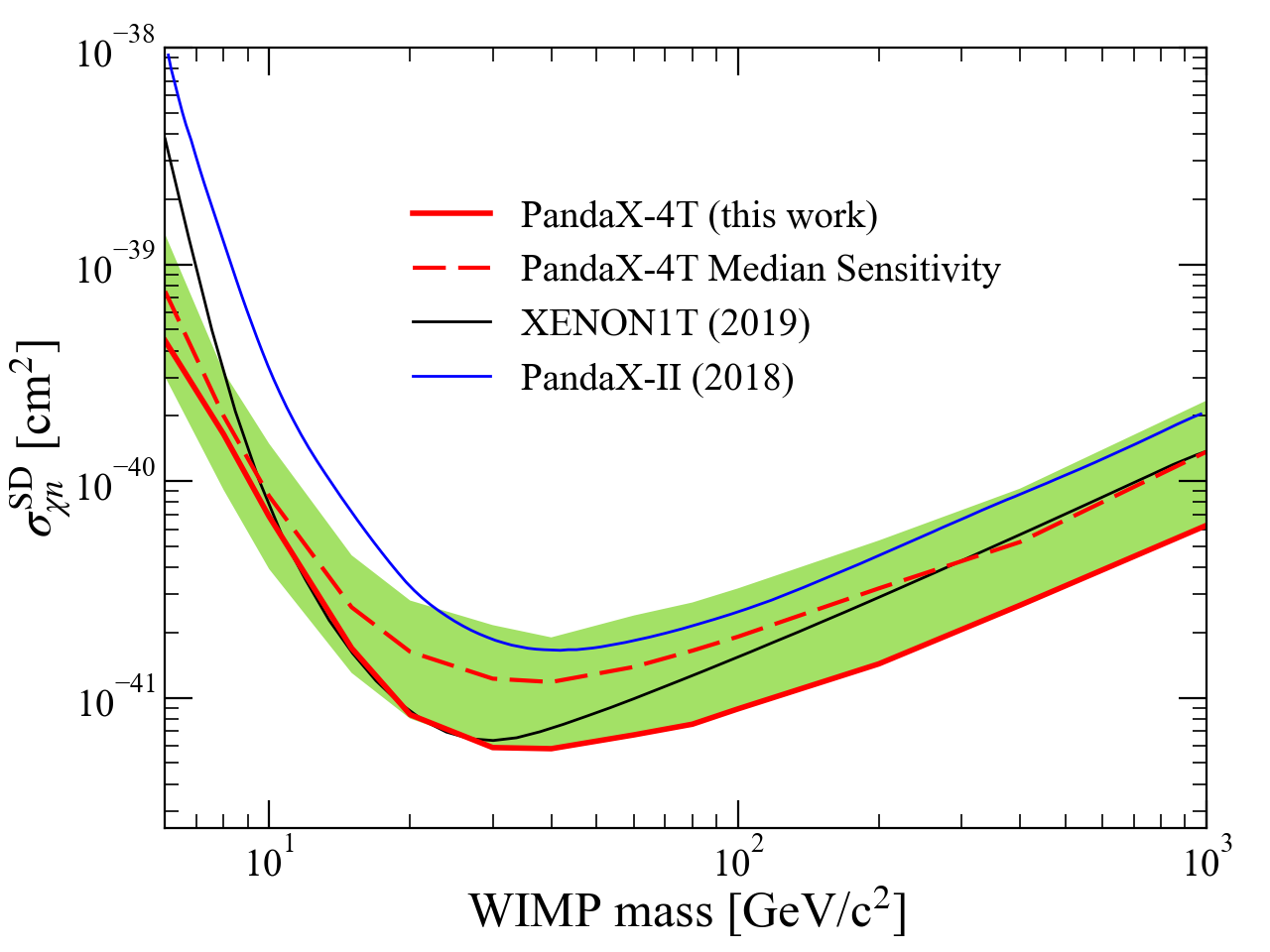}
\includegraphics[width=0.9\columnwidth]{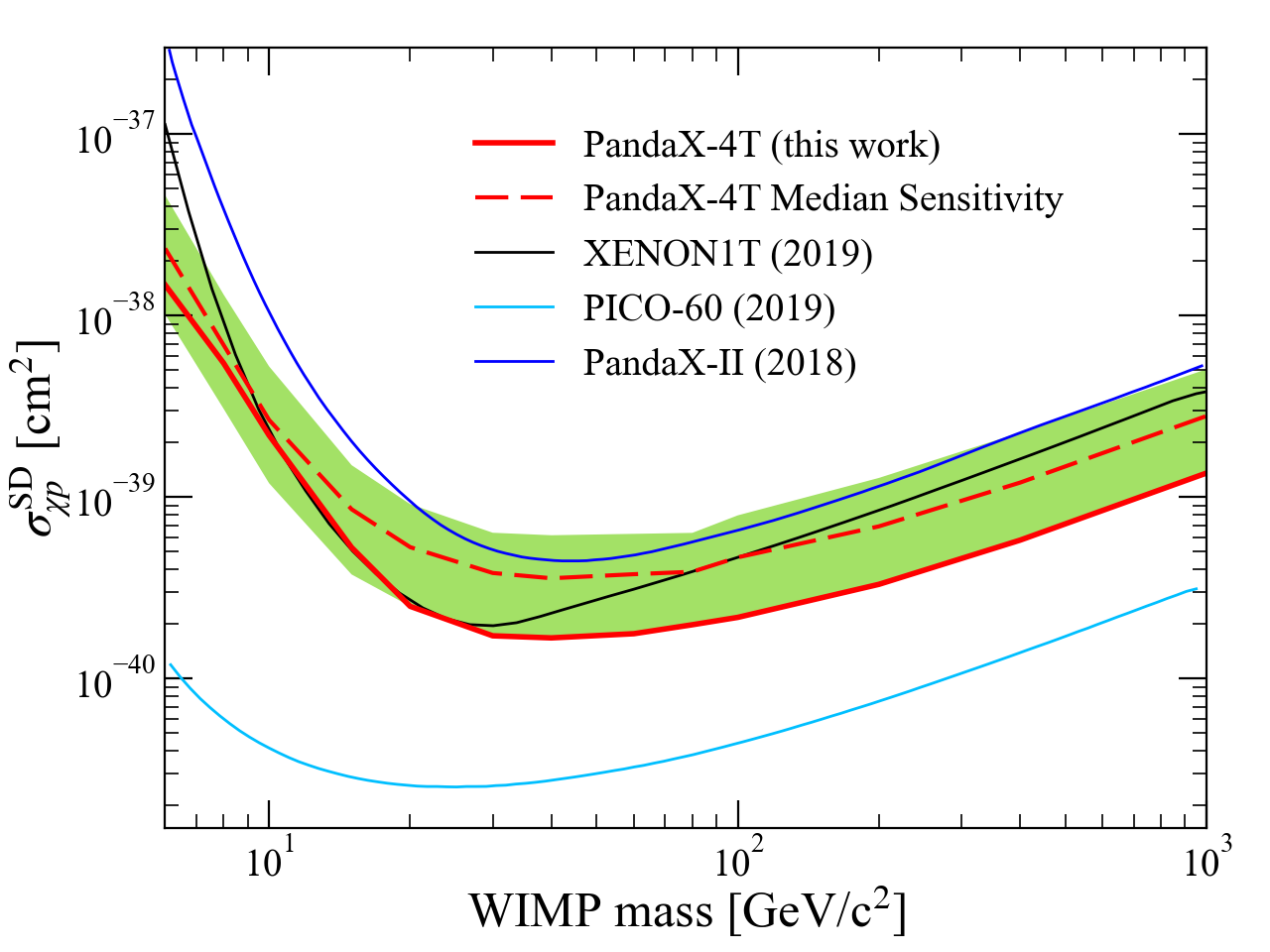}
\caption{
	PandaX-4T 90\% C.L. upper limit for the SD WIMP-nucleon (top: neutron-only; bottom: proton-only) cross-section,
	overlaid with that from XENON1T 2019~\cite{XENON1T-SD} and PandaX-II 2018~\cite{PandaX-II-EFT}.
    The green band and dashed red curve represent the $\pm~1\sigma$ sensitivity band and the median, respectively.
    }
\label{fig:limit_neutron}
\end{figure}

The 90\% confidence level (C.L.) upper limit of SD WIMP-nucleon cross-section as a function of WIMP mass is calculated,
utilizing the same statistical inference (profile likelihood ratio) in Ref.~\cite{PandaX-4T-SI, Baxter:2021pqo}.
The results are shown in Fig.~\ref{fig:limit_neutron}, % and Fig.~\ref{fig:limit_proton},
for ``neutron-only" and ``proton-only" interaction respectively.
Selected previous experiment results are also plotted for comparison. 
\textcolor{black}{In the WIMP mass above 6~GeV/c$^2$ region, our most stringent upper limit for "neutron-only" interaction is 5.8$\times$10$^{-42}$~cm$^2$
for a WIMP mass of 40~GeV/c$^2$. For the "proton-only" interaction, the minimal excluded cross-section is 1.7$\times$10$^{-40}$~cm$^2$ for 40~GeV/c$^2$ WIMP mass. Recently, the LZ experiment released its first result on the SD WIMP-nucleon interaction~\cite{LZ:2022ufs}, which extends the excluded SD parameter space further.}

%\begin{figure}[htbp]
%\centering
%\includegraphics[width=0.95\columnwidth]{figures/WIMP_proton_limit.png}
%\caption{	
%	PandaX-4T 90\% C.L. upper limit for the SD WIMP-proton cross-section,
%	overlaid with that from XENON1T 2019~\cite{XENON1T-SD}, PandaX-II 2018~\cite{PandaX-II-EFT} and PICO-60 2019~\cite{PICO:2019vsc}.
%	The green band and dashed red curve represent the $\pm~1\sigma$ sensitivity band and the median, respectively.
%	}
%\label{fig:limit_proton}
%\end{figure}

\begin{figure}[htbp]
\centering
\includegraphics[width=0.9\columnwidth]{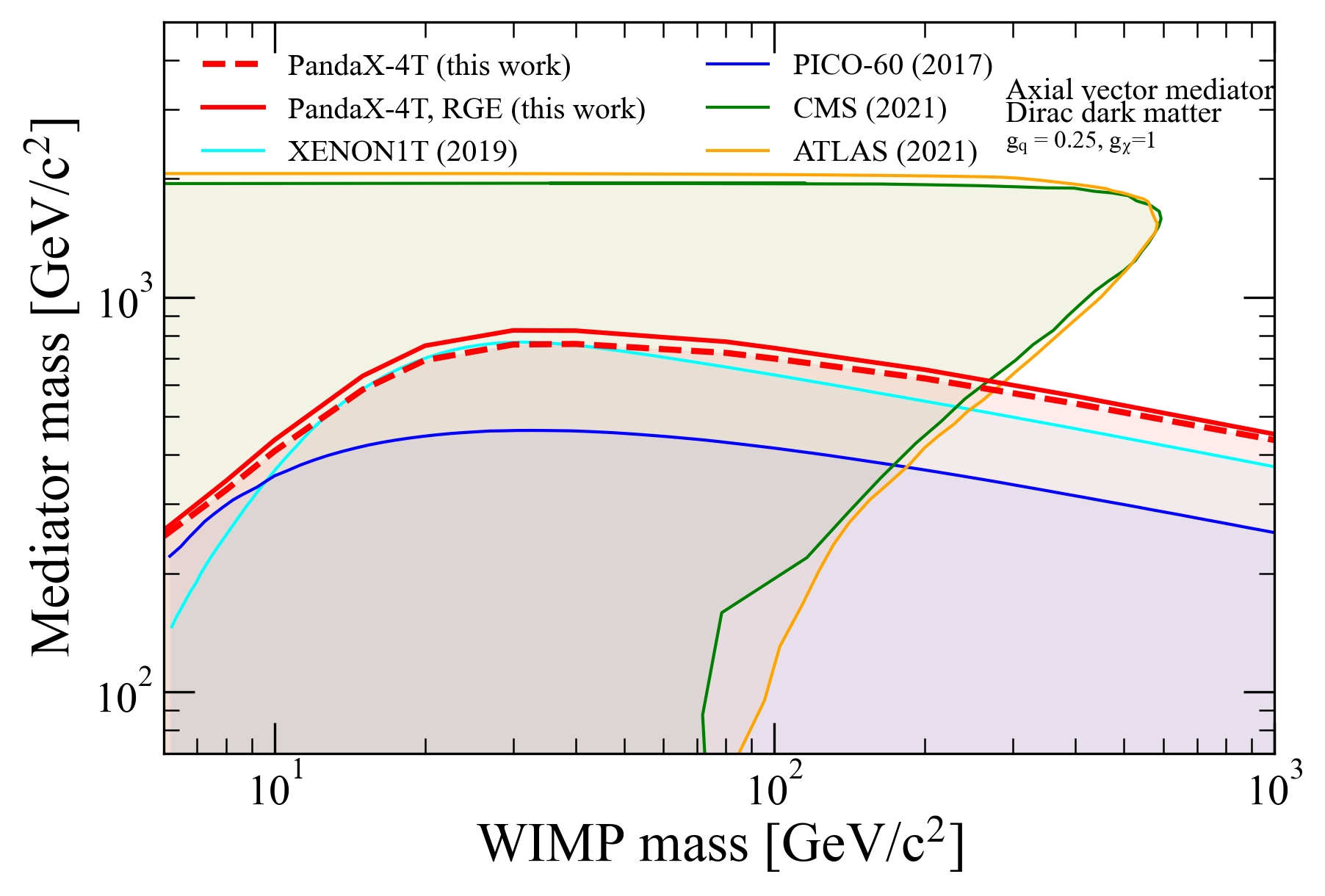}
\caption{PandaX-4T 95\% C.L. exclusion limit on the mediator mass 
for a simplified isoscalar model with an axial-vector mediator and a Dirac WIMP,
where the coupling constants $g_{ q}$ and $g_{\chi}$ are 0.25 and 1, respectively.
RGE effect is considered in the signal rate calculation, shown as a solid red line. The limit without RGE effect is shown as a dashed red line.
The results from other direct detection experiments, XENON1T~\cite{XENON1T-SD} and PICO-60~\cite{PICO:2017tgi},
and collider experiments, ATLAS~\cite{ATLAS:2021kxv} and CMS~\cite{CMS:2021far}, are overlaid for comparison.
}
\label{fig:limit_mediator}
\end{figure}

To explore the physics behind the WIMP-nucleon interaction, we consider the simplified model with a mediator. The axial-vector mediator couples to all quark flavors equally with one unique coupling constant $g_{q}$, and thus it yields isoscalar couplings to neutrons and protons.
In this model, there are four free parameters: WIMP mass $m_\chi$, mediator mass $m_V$, 
the coupling of the mediator to WIMP $g_\chi$ 
and the coupling of the mediator to quark $g_{q}$. 
Following the Ref.~\cite{Chao:2019lhb}, both the tree-level amplitudes and the loop-level amplitudes are calculated, and the latter is negligible.
The cross-section is given 
in Ref.~\cite{Boveia:2016mrp,Chao:2019lhb} with the form
\begin{equation}
	\sigma^{\rm SD}_{\chi N} = \frac{0.31 g_{q}^2 g_{\chi}^2 \mu_N^2}{\pi m_{\rm V}^4},
\end{equation}
where couplings $g_{ q} = 0.25$ and $g_{\chi} = 1$ are adopted as a benchmark scenario.
Based on this model, the signal energy spectra are generated and the same analysis method is applied. The 95\% C.L. exclusion limit on mediator mass as a function of WIMP mass is obtained, as shown in Fig.~\ref{fig:limit_mediator}, with a comparison with other direct detection experiments~\cite{PICO:2017tgi,XENON1T-SD}. In addition, the renormalization group evolution (RGE) effect is considered to account for the corrections from the mediator mass scale to the nuclear scale in direct detection~\cite{DEramo:2016gos}. The limit on the mediator mass with RGE effect is also shown in Fig.~\ref{fig:limit_mediator}. The maximum excluded mass of an axial-vector mediator is $825~\si{GeV}/c^2$. Constraints on the mediators are also provided by collider experiments such as the ATLAS and CMS at the Large Hadron Collider (LHC)~\cite{ATLAS:2021kxv,CMS:2021far}, with the benchmark models summarized in Refs.~\cite{LHCDarkMatterWorkingGroup:2018ufk,Boveia:2016mrp,Buckley:2014fba}.  This result is complementary to that obtained from the collider experiment in the WIMP mass region above several hundred GeV/c$^2$.

\begin{figure}
    \centering
  \includegraphics[width=0.95\columnwidth]{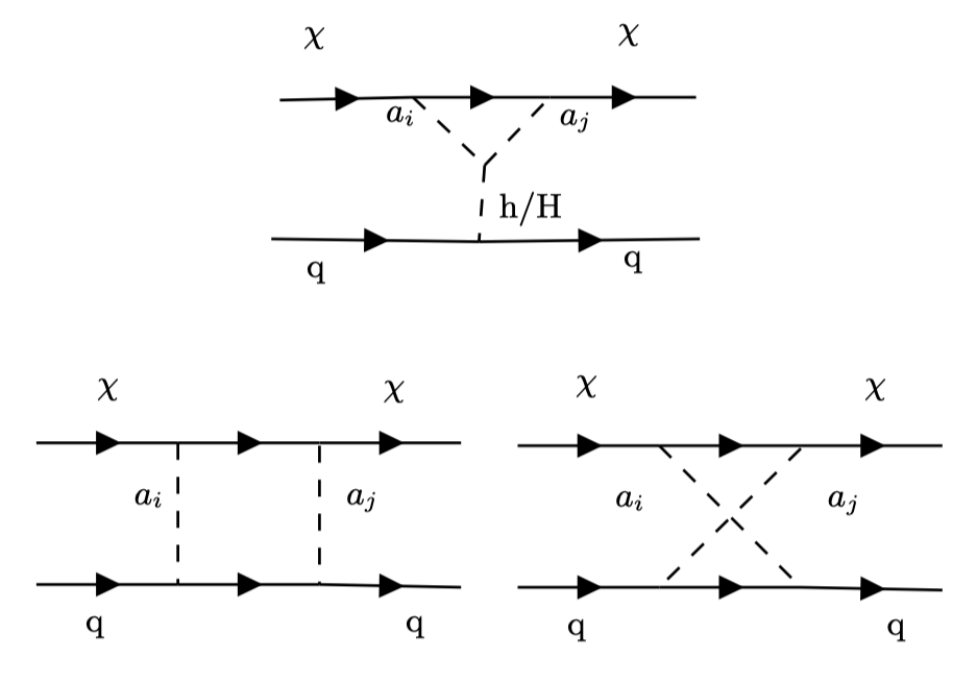}	
    \caption{The upper panel is the triangle diagram of scattering between WIMPs and quarks in 2HDM$+a$ model~\cite{abeLoopCorrectionsDark2019}, where $a_{i,j}$ represents the mediator $a/A$. The lower panel is the corresponding box diagrams in 2HDM$+a$ model~\cite{abeLoopCorrectionsDark2019}.}
    \label{fig:feynman}
\end{figure}

%\subsection{Pseudo-scalar}
\textcolor{black}{
As for the pseudo-scalar mediator scenario, one gauge-invariant and ultra-violet-complete extension of the simplified model is the Two-Higgs-Doublet-Model-Plus-Pseudo-Scalar (2HDM$+a$) model,
as recommended 
by the LHC DM Forum~\cite{LHCDarkMatterWorkingGroup:2018ufk}. 
It contains five Higgs bosons: a light CP-even boson, $h$, a heavier CP-even boson, $H$, a CP-odd boson, $A$, and two charged bosons, $H^{\pm}$, where the lighter CP-even boson $h$ can be identified with the Standard Model Higgs boson. A CP-odd pseudo-scalar mediator, $a$, is postulated with Yukawa-couplings to both the SM fermions and the Dirac fermion WIMPs, and mixes with the pseudo-scalar $A$ with a mixing angle $\theta$. 
However, with a pseudo-scalar mediator, 
the amplitudes of tree-level SD contribution are 
momentum-suppressed~\cite{Li:2018qip,Li:2019fnn}.
Therefore, the dominant processes are from the loop-level diagrams which give the SI interactions instead. The loop-level diagrams of interactions between WIMPs and quarks are shown as Fig.~\ref{fig:feynman}, and the amplitudes can be calculated from the effective Lagrangian following Ref.~\cite{Li:2019fnn,Bell:2018zra,abeLoopCorrectionsDark2019}. }

\textcolor{black}{This model contains 14 independent parameters: the masses of Higgs bosons $m_h$, $m_H$, $m_A$, and $m_{H^\pm}$; the mass of the mediator $m_a$; the WIMP mass $m_\chi$; the Yukawa coupling strength between the mediator and the WIMP, $g_\chi$; the electroweak vacuum expectation value (VEV), $v$; the ratio of the VEVs of the two Higgs doublets, $\tan\beta$; the mixing angles of the CP-even and CP-odd weak eigenstates, $\alpha$ and $\theta$, respectively; and the three quartic couplings between the scalar doublets and the mediator $(\lambda_{P1},~\lambda_{P2},~\lambda_3)$. In this letter, the following values of these parameters are adopted as suggested in Ref.~\cite{LHCDarkMatterWorkingGroup:2018ufk}:}
\begin{equation}
\begin{aligned}
    m_H &=  m_{H^{\pm}}  = m_A = 600~\si{GeV}/c^2,\\
    \cos(\beta-\alpha)&=0, ~\tan\beta=1, ~\sin\theta = 0.35, \\
    g_\chi& = 1, ~\lambda_3 = \lambda_{P1} = \lambda_{P2} = 3.
    \label{para}
\end{aligned}
\end{equation}

\begin{figure}[htbp]
\centering
\includegraphics[width=0.9\columnwidth]{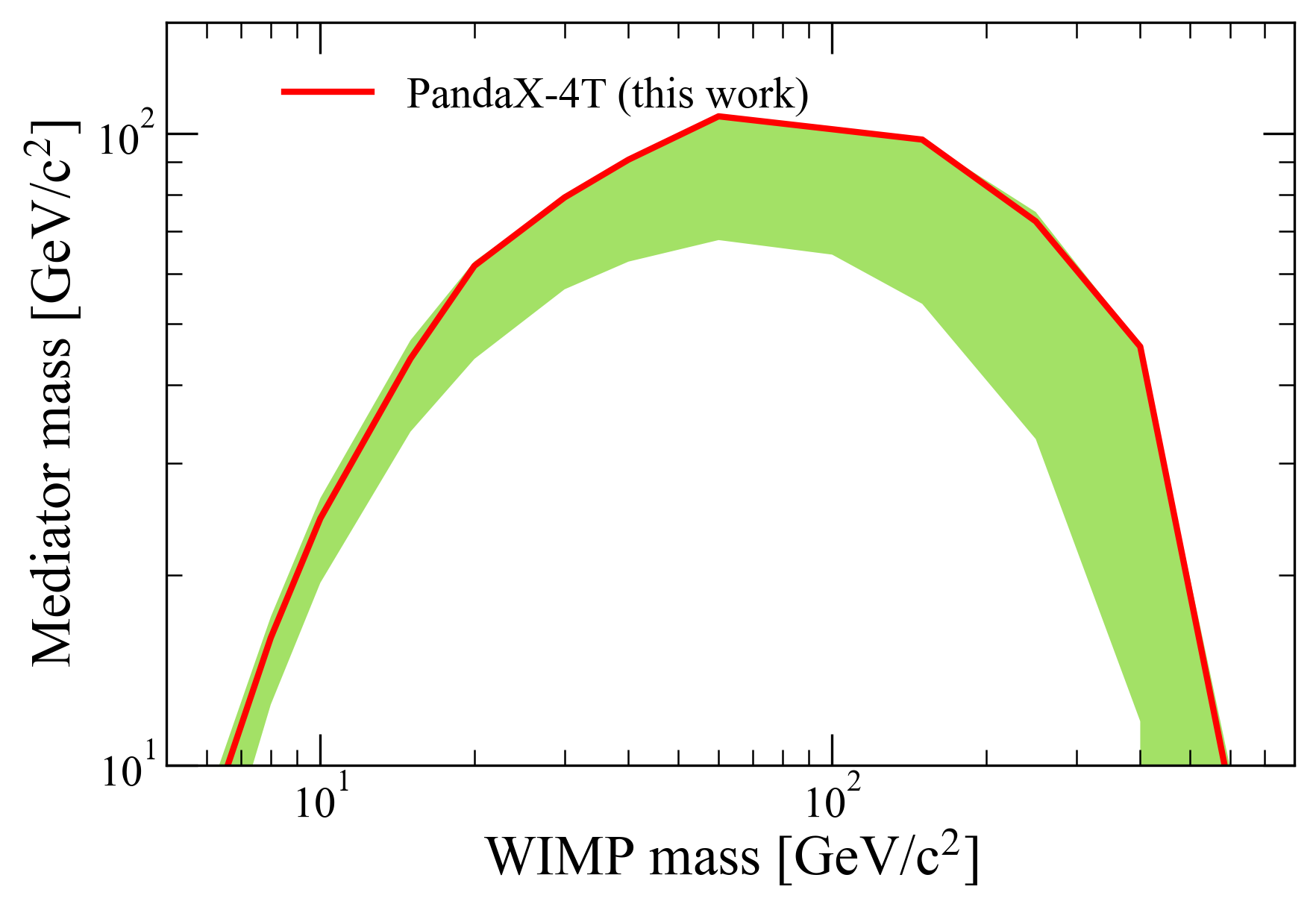}
\caption{PandaX-4T 95\% C.L. exclusion limit(red solid) on the mass of pseudo-scalar mediator $a$. The green band represents the $\pm 1 \sigma$ sensitivity band.}
\label{fig:ma}
\end{figure}

\textcolor{black}{With these parameters and signal rate calculation following Ref.~\cite{abeLoopCorrectionsDark2019}, the 95\% C.L. exclusion limit on the pseudo-scalar mediator mass $m_a$ as a function of the WIMP mass is shown in Fig.~\ref{fig:ma}, where the maximal excluded mediator mass is 106~GeV$/c^2$. Since the major contribution to the WIMP-nucleon scattering is from NLO processes in this scenario, the limits are much weaker than the axial-vector case, leaving a large parameter space viable for WIMPs. }

\begin{figure}
\centering
\includegraphics[width = 0.95\columnwidth]{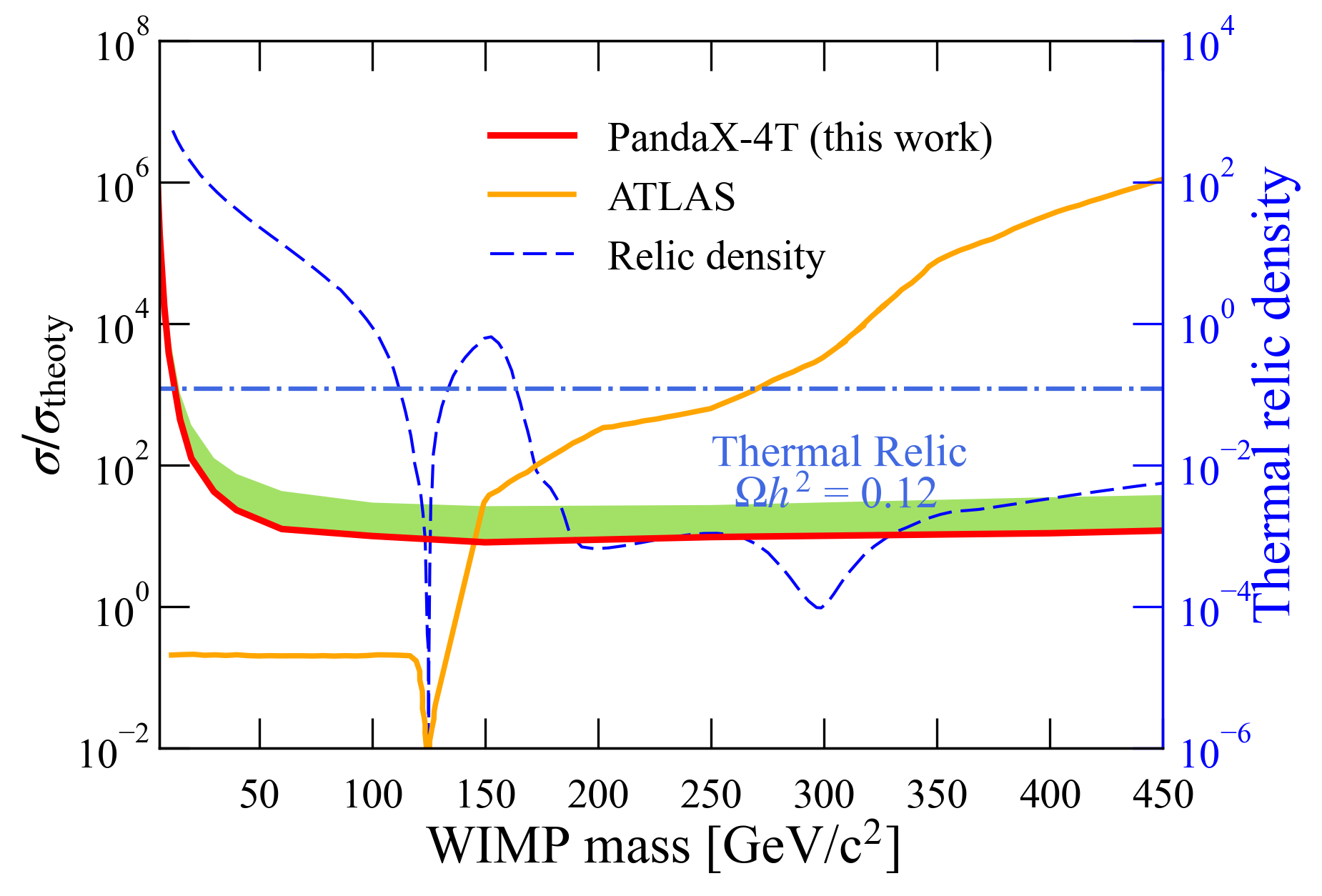}
\caption{
PandaX-4T 95\% C.L. normalized exclusion limit (red solid) for pseudo-scalar mediator model as a function of WIMP mass.
The green band shows the $\pm~1\sigma$ sensitivity band.
The normalized limit is expressed in terms of the ratio of the excluded cross-section
to the nominal cross-section of the model.
The 95\% C.L. from ATLAS ~\cite{ATLAS:2019wdu} (orange solid) is also shown for comparison.
The relic density (blue dashed) for each WIMP mass~\cite{ATLAS:2019wdu} is overlaid and described by the right blue axis, based on the parameters shown in~\eqref{para}.
}
\label{fig:limit_expectation}
\end{figure}

\textcolor{black}{PandaX-4T direct detection and collider search complement each other for this 2HDM$+a$ model, as illustrated in Fig.~\ref{fig:limit_expectation}, where the pseudo-scalar mass $m_a$ is fixed at $250~\si{GeV}/c^2$ as the collider search adopts.}
The exclusion limit is expressed in terms of the ratio of the excluded cross-section ($\sigma$) 
to the nominal cross-section of the model ($\sigma_{\rm theory}$) calculated with the benchmark parameters.
For $m_a=250~\si{GeV}/c^2$, the predicted thermal relic density as a function of WIMP mass in the 2HDM$+a$ model~\cite{ATLAS:2019wdu} is also shown in Fig.~\ref{fig:limit_expectation}, which indicates that only the WIMP mass range around 125~GeV$/c^2$ or above 180~GeV$/c^2$ is allowed.
Compared with the results from ATLAS experiment~\cite{ATLAS:2019wdu}, PandaX-4T experiment has much a stronger sensitivity to WIMPs with mass above 180~GeV$/c^2$.   

%\section{Summary and outlook}

\textcolor{black}{
In conclusion, this letter presents new constraints on the axial-vector and pseudo-scalar mediated WIMP-nucleus interactions from the PandaX-4T commissioning data of 0.63 tonne$\cdot$year. The most stringent constraints to date on the mediator mass are set at 90\% confidence level are set for the axial-vector and pseudo-scalar simplified models, taking into account both the tree-level and loop-level processes. The maximum excluded mass of an axial-vector mediator is $825~\si{GeV}/c^2$ and that of a pseudo-scalar mediator is $106~\si{GeV}/c^2$. 
%This result is complementary to that obtained from the collider experiment (ATLAS~\cite{ATLAS:2021kxv, ATLAS:2019wdu} and CMS~\cite{CMS:2021far})in the WIMP mass region above several hundred GeV/c$^2$.
Direct detection and collider search are
complementary to each other and the combination of these two directions covers a much larger parameter space of WIMPs and mediators.
In addition, the upper limits on the conventional spin-dependent WIMP-nucleon scattering cross-section are derived, with the minimum value of  $5.8 \times10^{-42}$~cm$^2$ and  $1.7 \times10^{-40}$~cm$^2$ for neutron-only and proton-only interactions respectively.}
The PandaX-4T experiment continues taking physics data and will scan unexplored parameter space further.

%\section{Acknowledgement}

We would like to thank Tong Li and Wei Chao for helpful discussion.
This project is supported in part by grants from National Science
Foundation of China (Nos. 1209061, 12005131, 11905128, 11925502, 11735003, 11775141), a grant from the Ministry of Science and Technology of China (No. 2016YFA0400301),
and by Office of Science and
Technology, Shanghai Municipal Government (grant No. 18JC1410200). We thank supports from Double First Class Plan of
the Shanghai Jiao Tong University. We also thank the sponsorship from the Chinese Academy of Sciences Center for Excellence in Particle
Physics (CCEPP), Hongwen Foundation in Hong Kong, Tencent
Foundation in China and Yangyang Development Fund. Finally, we thank the CJPL administration and
the Yalong River Hydropower Development Company Ltd. for
indispensable logistical support and other help.

\bibliographystyle{unsrt}
\bibliography{reference}

\end{document}